# Hybridization at the organic-metal interface: a surface-scientific analogue of Hückel's rule?†

Hasmik Harutyunyan,[a] Martin Callsen,[b] Tobias Allmers,[a] Vasile Caciuc,[b] Stefan Blügel,[b] Nicolae Atodiresei,[b§] and Daniel Wegner,[a‡]

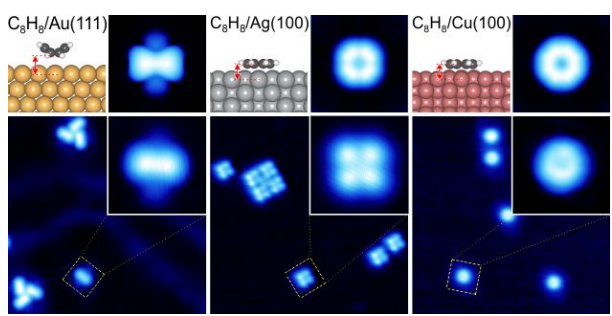

**We demonstrate that cyclooctatetraene (COT) can be stabilised in different conformations when adsorbed on different noble-metal surfaces due to varying molecule- substrate interaction. While at first glance the behaviour seems to be in accordance with Hückel's rule, a theoretical analysis reveals no significant charge transfer. The driving mechanism for the conformational change is hybridisation at the organic-metal interface and does not necessitate any charge transfer.**

† Electronic Supplementary Information (ESI) available: Experimental and theoretical methods, details on DFT and DFT-D3 results, additional experimental results, and acknowledgements. See DOI: 10.1039/b000000x/

[a] *Physikalisches Institut and Center for Nanotechnology (CeNTech), Westfälische Wilhelms-Universität Münster, 48149 Münster, Germany.*

[b] *Peter Grünberg Institut (PGI-1) and Institute for Advanced Simulation (IAS- 1), Forschungszentrum Jülich and JARA, 52425 Jülich, Germany.*

‡ Email: daniel.wegner@uni-muenster.de (experiment).

§ Email: n.atodiresei@fz-juelich.de (theory).

The fascinating prospect of molecular electronics is to contact and embed single molecules with chemically designed functions into electronic circuits.[1] A major challenge, however, remains to predict the changes of molecular properties at the organic-metal interface. Charge transfer, hybridisation and screening effects can dramatically modify the molecular properties and may even lead to a loss of functionality with respect to molecules in the gas phase.[2] Charge-transfer and hybridisation processes are often discussed independently. The former can be connected to molecular orbitals crossing the Fermi energy and thus changing their occupation. The latter is reflected in the energetic broadening of orbitals and the formation of new electronic states with mixed molecular and metallic character. However, in cases of strong hybridisation, charge transfer is not necessarily well-defined anymore, raising questions of comparability with the gas-phase molecule.

In this context, the annulene 1,3,5,7-cyclooctatetraene (COT, $C_8H_8$) may offer interesting insights.[3] According to Hückel's rule, a planar organic ring molecule is aromatic when ($4n + 2$) π-electrons are present ($n = 0, 1, 2, ...$). Neutral COT in its ground state is found in a non-aromatic tub conformation ($D_{2d}$ symmetry),[4] since it contains eight π-electrons and would be antiaromatic when planar. When COT is doubly charged, Hückel's rule is fulfilled, and the dianion is planar and aromatic ($D_{8h}$ symmetry).[5] Already the singly charged anion is planar, but due to localised alternating single and double bonds ($D_{4h}$ symmetry) it is not aromatic.[6] Thus, COT is an excellent candidate for studying molecule-surface hybridisation effects: the actual conformation of COT adsorbed on a metal surface can be probed in real space, while a theoretical analysis can shed light on the actual charge transfer and hybridisation.

By combining low-temperature scanning tunneling microscopy (STM) and density functional theory (DFT), we investigated COT adsorption on noble-metal surfaces with different degrees of hybridisation (for experimental and theoretical details, cf. Supplementary Information[†]). We first discuss STM results for COT deposition on Au(111). As seen in Fig. 1(a,b), the molecules have formed loosely bound small clusters. For tunneling currents exceeding 80 pA, molecules could be moved with the STM tip. A close-up view of a single COT (inset in Fig. 1b) reveals that the molecular shape consists of a bright elongated protrusion with an apparent height of ~1.6 Å and two faint protrusions in perpendicular direction, independent of the tunneling parameters. Small clusters arrange close-packed, but a disordered arrangement is observed in larger clusters.

On Ag(100), the STM images reveal that COT molecules remain mainly isolated (Fig. 1f,g), but few small clusters are still visible. Manipulation experiments showed that single molecules can only be moved along the surface using relatively large tunneling conductances (~10 mV, 10 nA). Highly resolved STM images at close tip-sample distances show that each molecule exhibits four lobes in square arrangement (average apparent height ~0.7 Å) with node lines oriented along the [100] directions (g).

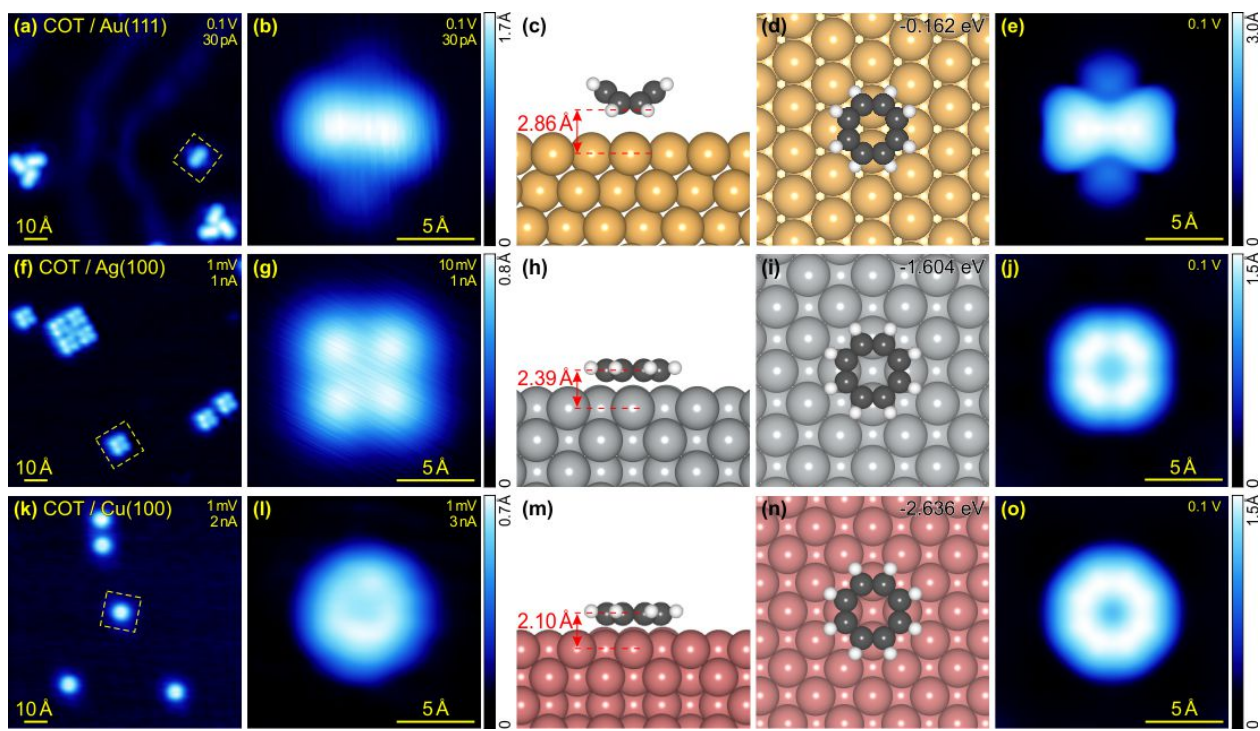

**Fig. 1** STM images and DFT results of COT deposited on Au(111) (top row), Ag(100) (middle), and Cu(100) (bottom). At positive bias, electrons tunnel from tip to sample.[†] STM images indicate increasing molecule-substrate interaction, respectively. Highly resolved images of single molecules reveal twofold, fourfold, and rotational symmetry, respectively. DFT results and simulated STM images are in good agreement with experiments, showing a decreasing molecule-surface distance. On Au(111), the tub conformation is most stable, while a flat conformation is found on Ag(100) and Cu(100). Adsorption energies (d,i,n) are obtained without vdW corrections.[†]

Finally, we deposited COT onto the more reactive surface Cu(100).[7] In this case, STM images (Fig. 1k,l) show that all molecules remain isolated. Furthermore, lateral manipulation was only possible for rather large tunneling conductances (~1 mV, 10 nA). Each molecule appears as a rotationally symmetric ring-like protrusion, with a faint depression at the molecular centre (cf. inset of Fig. 1l) and apparent height identical to that observed on Ag(100).

A direct experimental comparison of COT adsorption already demonstrates an increasing molecule-surface interaction as we go from Au to Ag to Cu, respectively. The change of molecular symmetries is indicative of different conformational states. In order to assign the respective molecular geometries and electronic properties unambiguously, we have performed DFT calculations.[†] On Au(111), COT adsorbs on bridge sites in the tub conformation (Fig. 1c,d). The simulated STM image (e) reveals that the topography is dominated by the two arms of the octagon that are tilted away from the surface. They produce an elongated protrusion waisted at the molecular centre, while the two arms tilted toward the surface produce faint protrusions in perpendicular direction. This is in very good agreement with the experimental STM topography. Regarding the bonding mechanism, our study predicts a DFT binding energy of -0.162 eV and a molecule-surface distance of 2.86 Å, which corresponds to physisorption. When including van der Waals (vdW) interactions,[8] the adsorption energy is lowered to -0.68 eV while the geometry does not change significantly.[†] The molecular conformation corresponds to that of the neutral non-aromatic molecule, we only find an increased bond angle within the ring compared to the free molecule. Weak interaction of COT with the Au surface is also apparent from the analysis of the projected density of states (PDOS) (Fig. 2b, top row): at the molecular site, the sharp PDOS peaks are similar to those of the molecule in the gas phase (Fig. 2a). Thus, the rather weak bond is in accordance with the experimental findings.

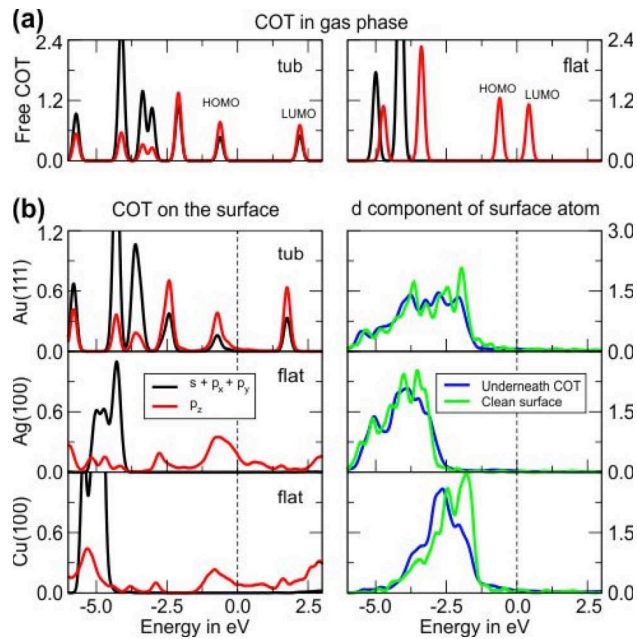

**Fig. 2** Calculated PDOS for neutral tub and flat COT in the gas phase (a), and for COT on noble metals (b). The degree of hybridisation reveals physisorption on Au(111) and increasing chemisorption on Ag(100) and Cu(100), respectively.

The DFT results of COT on Ag(100) reveal that the molecule adsorbs on hollow sites and has a planar conformation with $D_{4h}$ symmetry (Fig. 1h,i). The C=C double bonds are those in horizontal and vertical directions (cf. Table S1).[†] The DFT binding energy is -1.604 eV, and the molecule-substrate distance is reduced to 2.39 Å. Therefore, the planar conformation can be understood by the onset of hybridisation of the molecular and substrate electronic states. This structural change corresponds to the transition from physisorption on Au(111) to chemisorption on Ag(100). As seen in the PDOS (Fig.2b, middle row), a strong broadening of the electronic features at the molecular site is a clear indication of the hybridised molecule-substrate states. The simulated STM image (Fig. 1j) is again in good agreement with the experiments and confirms that the electronic molecular shape of the hybrid orbitals in the relevant energy range mostly resembles the HOMO shape of the free molecule.

COT on Cu(100) also adsorbs on hollow sites and shows a planar conformation (Fig. 1m,n). However, the DFT binding energy is -2.636 eV and the molecule-substrate distance is 2.10 Å, i.e., the molecule is chemisorbed much stronger on Cu(100). This is also reflected in the PDOS (Fig. 2b, bottom row). Interestingly, our simulated STM image (Fig. 1o) confirms the experimentally observed ring-like shape, although the geometrical structure of COT on Cu(100) has a $D_{4h}$ symmetry, as on Ag(100). Nevertheless, the stronger hybridisation to the Cu(100) substrate causes a local increase of the charge density in the molecular π-channel, leading to an apparent rotational symmetry in STM imaging.

On the one hand, the resemblance between the experimentally observed molecular shapes and symmetries on the three surfaces and those known for the gas-phase species in different charge states is striking. The physisorbed tub conformation on Au(111), a surface generally known for relatively weak interactions with organic molecules,[7,9] is in analogy to non-aromatic neutral $COT^0$. For Ag(100) and Cu(100), an increased interaction leads to planarisation of the chemisorbed COT with fourfold symmetry on Ag(100) analogous to the anion $COT^{1-}$ and a ring-like shape with rotational symmetry on Cu(100) similar to the aromatic anion $COT^{2-}$. On the other hand, our analysis of calculated charge distributions shows no significant charge transfer at the molecular site on all three surfaces, which forbids such a simple explanation. This clearly demonstrates that the charge state of COT on noble metals is not well-defined anymore. Besides, molecular adsorption can lead to formation of new hybrid states that neither exist in the free molecule nor on the bare surface.[9,10] Such a scenario renders the attribution of charge density even more impossible. Hence, in the present case the modification of the molecule-surface interaction through the choice of metal should be considered a hybridisation-driven mechanism that can obviously have a comparable effect as charging of the molecule in the gas phase.

The impact of varied degree of hybridisation on the COT conformation can even be observed directly on the Au(111) surface. The herringbone reconstruction exhibits local reactive sites due to point dislocations (corners of the discommensuration lines).[11] Indeed, we found that COT appears as round protrusion at these sites, revealing a planar molecular conformation (for more details, see Supplementary Information[†]).

In summary, our results demonstrate that we can identify the molecular conformation of COT via altering its hybridisation to a metallic substrate. The resemblance between our observations for the adsorbed molecule and the expected con- formations of the gas-phase molecule in different charge states is striking. However, we showed that due to the lack of well-defined charge states for adsorbed molecules, the conformational change is hybridisation-driven. In this respect, Hückel's rule is not applicable for chemisorbed molecules.

Supplementary Information for

# Hybridization at the organic-metal interface: a surface-scientific analogue of Hückel's rule?

Hasmik Harutyunyan,[a] Martin Callsen,[b] Tobias Allmers,[a] Vasile Caciuc,[b] Stefan Blügel,[b] Nicolae Atodiresei,[b] and Daniel Wegner[a‡]

[a] *Physikalisches Institut & Center for Nanotechnology (CeNTech), Westfälische Wilhelms-Universität Münster, Wilhelm-Klemm-Str. 10, 48149 Münster (Germany)*

[b] *Peter Grünberg Institut (PGI-1) and Institute for Advanced Simulation (IAS-1), Forschungszentrum Jülich and JARA, 52425 Jülich (Germany)*

[§] *Email: n.atodiresei@fz-juelich.de (theory)*

[‡] *Email: daniel.wegner@uni-muenster.de (experiment)*

**Experimental Methods**

The experiments were performed in ultrahigh vacuum (UHV) using a commercial low-temperature STM (Createc LT-STM) operated at 5.5 K. The Au(111), Ag(100) and Cu(100) single-crystal substrates were cleaned by standard sputter-annealing procedures and then transferred *in situ* into the cold STM. Owing to its high vapour pressure, COT (Sigma Aldrich, 98% purity) was injected into the UHV system as a pure gas through a leak valve. The molecules were deposited onto the cold sample inside the STM by opening the cryostat shutter for about 60 seconds. During this process, the partial pressure was about $10^{-8}$ to $10^{-7}$ mbar and the sample temperature always stayed below 20 K. The bias voltage in STM acquisition is referred to the sample, i.e., at positive (negative) voltages electrons tunnel from the tip (sample) to the sample (tip) thus probing the unoccupied (occupied) local density of states of the sample.

**Theoretical Methods**

Density functional theory (DFT) calculations were performed with the VASP code using the projector augmented wave (PAW) pseudopotential method with a plane-wave cut-off energy of 500 eV.[S1–S4] The exchange correlation (xc) effects were described by the Perdew-Burke-Ernzerhof (PBE) functional.[S5] The influence of long-range dispersion interactions, that are not correctly taken into account by the local or semi-local exchange-correlation functionals,[8] has been investigated using the semi-empirical DFT-D3 approach.[S6] For all three surfaces, the molecule-substrate system was modelled in a slab geometry consisting of six metal layers with the molecule adsorbed on one side of the slab. The ground state was obtained by relaxing the molecule as well as the upper three metal layers until forces were smaller than 0.002 eV/Å. STM images have been simulated within the Tersoff-Hamann model.[S7]

**Details of DFT calculations**

Table S1 summarizes calculated bond lengths, molecule-surface distances, and adsorption energies for COT on the three noble-metal surfaces. The binding (adsorption) energy $E_{ads}$ is defined as the difference between the total energy of the relaxed molecule-surface system $E_{sys}$ and those of the molecule in the gas phase $E_{molec}$ and the clean surface $E_{surf}$, i.e., $E_{ads} = E_{sys} - (E_{molec} + E_{surf})$.

**Table S1.** Calculated C–C single and C=C double bond lengths as well as DFT and DFT-D3 molecule-surface equilibrium distances ($d_{eq}$) and corresponding adsorption energies $E_{ads}$ of COT on Au(111), Ag(100), and Cu(100). $d_{eq}$ is defined as the distance between plane spanned by the substrate's surface atoms underneath the molecules and that spanned by the lowest molecular atoms (which is hydrogens in case of Au and carbons in case of Ag and Cu).

| Surface | C–C (Å) | C=C (Å) | $d_{eq}$ (Å) | | $E_{ads}$ (eV) | |
|---|---|---|---|---|---|---|
| | | | DFT | DFT-D3 | DFT | DFT-D3 |
| Au(111) | 1.474 | 1.349 | 2.86 | 2.71 | -0.162 | -0.679 |
| Ag(100) | 1.436 | 1.416 | 2.39 | 2.39 | -1.614 | -2.174 |
| Cu(100) | 1.443 | 1.418 | 2.10 | 2.10 | -2.636 | -3.389 |

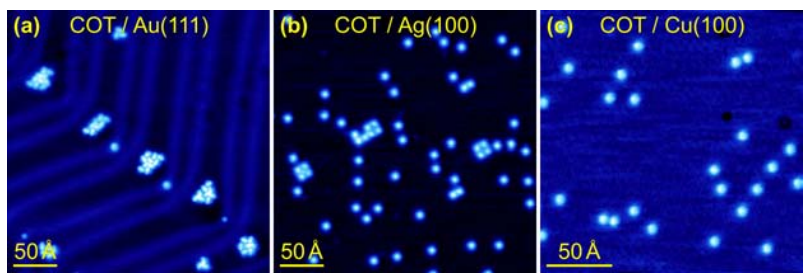

**Figure S1.** Overview images of COT adsorption on the three different noble-metal surfaces. Imaging parameters: (a) 1 V, 90 pA; (b) 1 V, 160 pA; (c) 1 mV, 1 nA

## STM overview images

Fig. S1 displays large-scale images of COT on Au(111), Ag(100), and Cu(100), respectively. On Au(111), all molecules are clustered, despite the low deposition temperature of $T_{dep} < 20$ K. For the study of a single COT, we pulled out molecules from a cluster via lateral STM manipulation.[S8] On Ag(100), only few clusters with a c(4x4) arrangement are found on Ag(100), while most molecules are isolated. On Cu(100), all COT molecules are isolated, i.e., no cluster has been found anywhere on the surface.

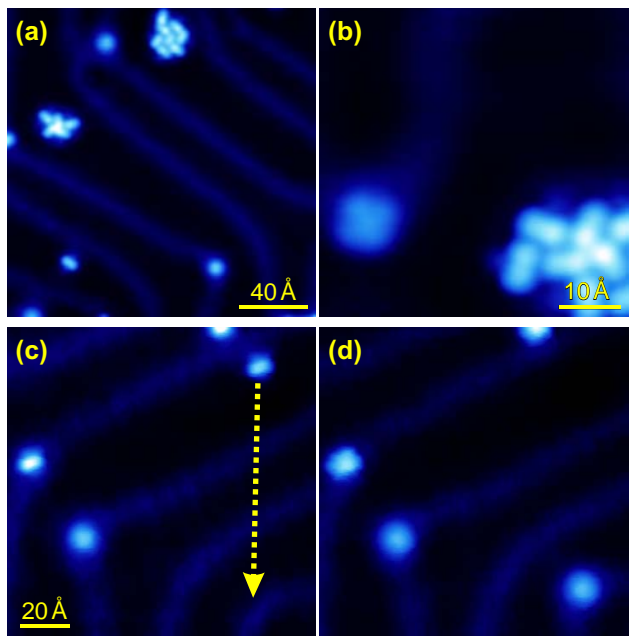

**Figure S2.** Lateral control of COT conformation on Au(111). (a) COT molecules adsorbed at elbow dislocations of the herringbone reconstruction exhibit a flat round protrusion different from the tub conformation. (b) A highly-resolved image of COT at an elbow dislocation (left) reveals a cross-like shape indicative of a planar conformation. (c) A tub-shaped COT is moved to an exposed elbow dislocation (arrow) (d) The molecule has changed to a planar conformation. Tunneling parameters: (a) 1 V, 80 pA; (b) 0.36 V, 0.4 nA; (c,d) 0.5 V, 1 pA.

**Further results on Au(111)**

In order to corroborate the general concept that the degree of hybridization determines the COT conformation, we can directly compare its properties on a surface with locally varying reactivity. Here we make use of the Au(111) herringbone reconstruction that can be seen as faint stripes in all our STM images on Au(111). The elbows (bends of the stripes) of the so-called *x*-type discommensuration lines exhibit point dislocations that are more reactive than other parts of the surface, often serving as nucleation sites for atoms and molecules.[11,S9–S12] Indeed, we also found that initial adsorption of COT on Au(111) starts by occupying the dislocation sites with a single molecule, respectively (cf. Fig. S1(a) and Fig. S2). Each COT appears as round protrusion with an apparent height of about 1.0 Å, i.e., much lower than the elongated COT on other parts of the Au surface. Highly resolved images reveal a cross-like pattern with fourfold symmetry (Fig. S2(b)) reminiscent of COT on Ag(100). These observations are clear evidence for a change of molecular conformation from tub to planar geometry upon adsorption of COT at Au(111) dislocation sites.

Going a step further, the conformation of COT can even be altered by means of STM manipulation.[S8] Fig. S2(c) shows an STM image with two COT molecules in the tub conformation (left side and top right corner), a molecule in planar conformation adsorbed at an elbow dislocation (bottom left), and a free elbow site (bottom right). As indicated by the arrow in (c), one of the tub-shaped COT was moved to the dislocation (typical manipulation parameters: 50 mV, 0.5 nA). The image after this manipulation (Fig. S2(d)) shows that the molecule is now adsorbed at the dislocation and has changed its shape. It is now identical to that of the other elbow-adsorbed COT, i.e., it has changed from tub to planar conformation.

**Bias dependence of apparent heights**

Fig. S3 summarizes the apparent height of COT molecules on the three noble metals as a function of sample bias. There is virtually no bias dependence for Au(111). The apparent heights on Ag(100) and Cu(100) behave almost identical, and we observed a slight decrease for increasing sample voltages. The lack of significant bias dependent variations in the apparent height indicate that no sharp molecular resonances are present in the observed energy window ±1 eV around $E_F$. Also simulated STM images show no significant voltage dependence. We note that for a particular set bias measured heights vary by several 0.1 Å. This is due to different set-point currents but also due to the influence of different STM tips.

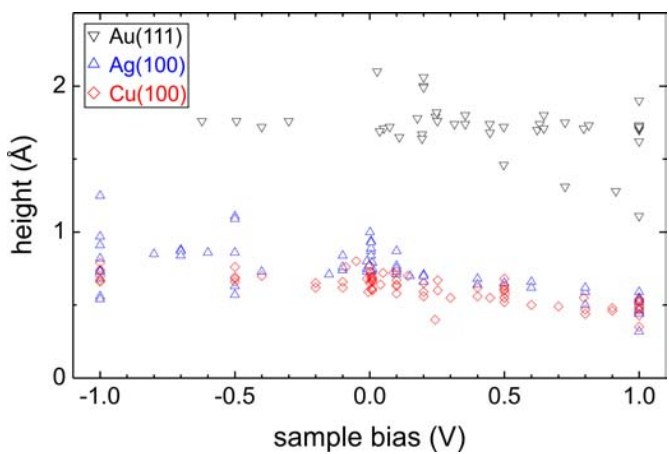

**Figure S3.** Apparent height vs. sample bias of COT molecules on Au(111), Ag(100), and Cu(100). The observed bias dependence is relatively weak.

**Acknowledgements**

We thank the Deutsche Forschungsgemeinschaft (DFG) for financial support through projects WE 4104/2-1 and SPP 1243. Computations were performed under the auspices of the VSR at the computer JUROPA and the Gauss Centre for Supercomputing at the high-performance computer JUGENE operated by the Jülich Supercomputer Centre at the Forschungszentrum Jülich.